# Design of a Community-based Translation Center

**Kathleen McDevitt**
Dept. of Computer Science
Virginia Tech
+1 540 552 2709
kmcdevit@vt.edu

**Manuel A. Pérez-Quiñones**
Dept. of Computer Science
Virginia Tech
+1 540 231 2646
perez@cs.vt.edu

**Olga I. Padilla-Falto**
Dept. of Foreign Languages
and Literature
Virginia Tech
+1 540 231 9846
padillao@vt.edu

**ABSTRACT**
Interfaces that support multi-lingual content can reach a broader community. We wish to extend the reach of CITIDEL, a digital library for computing education materials, to support multiple languages. By doing so, we hope that it will increase the number of users, and in turn the number of resources. This paper discusses three approaches to translation (automated translation, developer-based, and community-based), and a brief evaluation of these approaches. It proposes a design for an online community translation center where volunteers help translate interface components and educational materials available in CITIDEL.

**Keywords**
translation systems, internationalization, localization, user interface, community-computing, task-based community, community-based transition, digital library

**INTRODUCTION**
Providing a multi-lingual interface and content is a good way to expand the user population of a web-based system, especially as the internet and computing fields grow more and more diverse. There are many ways of providing translations, however some yield better translations than others. This paper presents a discussion of several approaches to translation of content for an interface as well as a brief evaluation of these approaches. It then concludes with the design of a voluntary community-based collaboration center that we expect will help produce quality translations of the interface of a digital library for computing education.

An existing digital library, The Computing and Information Technology Interactive Digital Education Library (CITIDEL) [3], serves the educational computing community with helpful resources and tools. This online digital library can be used by students, professors, researchers and professionals alike as a portal to a number of collections of computing resources. By providing an interface for CITIDEL in multiple languages, we hope to not only increase use and numbers of users, but also the number of resources themselves.

**BACKGROUND**
As the internet allows information to pass easily around the world, many companies and organizations wish to participate in this globalization. Some select specialized companies to translate their content for them, while others do it themselves or create special international branches and delegate this work. There are a number of different approaches to translating a website, each with its own pros and cons.

Qualities to consider for translation of an interface include: quality of resulting translation and industry-specific terms, style of resulting translation, party in control, time required of those running the site, user reviewing, dispute resolution, and flexibility and layout of the interface.

The *quality of a translation* is greatly dependant upon the translators' skills with the languages. It also includes such aspects as accurate mappings, handling abbreviations, and proper grammar. *Industry-specific terms*, such as computer terms, may or may not have accurate mappings between languages, or might have multiple ones, depending upon connotation. *Style* of the translation is important for large sites or documents where certain terms must be referred to consistently, and when the tone of a site must be maintained. *Control* is an important issue, as many developers are reluctant for security reasons to allow decisions about the interface to be made by individuals outside the organization.

As sites grow and change, translations must reflect this, and for administrators who run a site to devote their time to translating or fixing errors results in less time devoted to other issues such as the content of the site. Furthermore, most site designers or administrators might not have the require knowledge to produce a website in multiple languages.

In many cases, it is not simply enough to translate words into another language. Languages with scripting systems that read from right to left, for example, should have layouts and menus adjusted accordingly when translated from English. Furthremore, Marcus and Gould suggest that, based on Hofstede's five dimensions of culture, interfaces for countries of different cultures should look and feel very different [7]. These cultural dimensions can influence aspects of the interface from the way information is displayed to the spacing between sections to the colors used.

Jakob Nielsen has written and spoken extensively on the subject of designing interfaces for international use [6, 10, 11]. He is a proponent of optimizing a single design for an interface wherever possible. However, as he explains, translating an interface is much different than simply translating the text, and many usability principles and cultural issues must be taken into account. Quality translations are essential, and as is quality design, Nielsen recommends some usability testing of translated interfaces as well.

To successfully cross the language translation divide, Dilts [4] presents many suggestions for both writing clear, translatable content as well as choosing a suitable translation service. Among these, are leaving proper space for translations in languages which require more space, and providing translators with definitions of computer industry jargon.

The choice of system used for translation is personal and all of the characteristics described above should be considered when selecting an appropriate way to translate a site.

**Automated Translation**
There are a number of ways to go about providing translations of interfaces. One such way is translating interfaces with an automated translation program, though quality of these results can vary greatly. An example of such an automatic system is AltaVista's Babel Fish Translation [1] in which users may receive translations for words, sentences, or paragraphs by entering these directly, or translating a full web page by providing a url to a page. Babel Fish is based on a product by Systran [17].

Currently, Babel Fish Translation supports translations from English to eight different languages and from eleven different languages into English. Another feature of the system allows website developers to add a box to their site so that users may use translate their website or parts of it remotely (on the fly). Automated translation systems such as these are quick and easy solutions, but their preprogrammed vocabulary and rules yields rough translations which can be inaccurate or fail to take slang or industry vocabulary into account. Other problems occur with page layouts that may change drastically because of space constraints in the translated language or with images containing words which are not translated. The only flexibility offered is to accept or reject the translations, and there is no ability to edit or make corrections in the translated product.

**Developer/Administrator-Based**
Another approach is for the designer or administrator of a site to translate the interface by hand, which is a cumbersome and time-consuming task for developers. The problems which can arise in this approach are discussed well in "Internationalizing Online Information" by Merrill and Shanoski [8]. Many times, developers ignore internationalization issues when designing a site and must reconsider these at the time of deployment.

Issues such as leaving enough space for languages which will have more characters to display, not using abbreviations or properly compensating for them, acknowledging numbers and currency, and considering culturally-friendly colors, layouts and symbols are left up to those in charge of running the site.

This takes time away from other duties of the developers' and administrators'. It also puts the responsibility of the translating on a set of finite individuals, with the quality depending upon these few people, and making future editing more difficult. Nonetheless, this is a common solution taken when localizing or internationalizing websites.

**Community-Based**
A third option is appeal to a site's user-base and allow this select community of volunteers to translate the interface. This approach has been successful in translating online newsletters and entire websites (more details are given below). This approach depends on an online community of users to volunteer their time to do the translation.

There are many different definitions of what an online community is. Jenny Preece in her book *Online Communities, Designing usability and supporting sociability* [15] defines an online community as a place online that has:

- "People, who interact socially as they strive to satisfy their own needs or perform special roles, such as leading or moderating.
- A shared purpose, such as an interest, need, information exchange, or service that provides a reason for the community.
- Policies, in the form of tacit assumptions, rituals, protocols, rules, and laws that guide people's interactions.
- Computer systems, to support and mediate social interaction and facilitate a sense of togetherness."

A task-based community is an online community formed around a central task, such as the task to translate an interface. To allow a task-based community to translate portions of a site's interface requires a bit of faith, but can yield quality results. The community of users of the system has an invested interest in it, and thus it is natural to appeal to them to translate. Members of the community can serve as a system of checks and balances as well as a consensus when there are disputes. There are concerns with this approach such as security in giving the responsibility of translations (content) to users and dispute resolution when members of the community disagree on particular translations. But often the practical benefits outweigh these drawbacks. By providing tools and ways for users to collaborate, translations of consistent quality and style are much more likely.

|  | **Automated system** | **Developer/Administrator** | **Community-based** |
|---|---|---|---|
| **Quality of translation** | Depends on particular program | Depends on human talent | Depends on collaboration among volunteers |
| **Quality of technical terms** | Good only when specifically programmed | Depends on human talent | Depends on collaboration among volunteers |
| **Style of translation** | Poor | Great | Good/Great (depends on collaboration) |
| **Control of translation** | Pre-programmed system (computer) | Developer/Administrator (human) | Community (humans) |
| **Developer's time required** | Little | Much | Little |
| **User review system** | Ability to accept/reject | Requires evaluation to identify errors | Possible ability to rate and write reviews among community |
| **Resolution of disputes** | Not needed | Resolved by developers/administrators on their on discretion | Resolved by community members in reviews or procedures or by administrators |
| **Flexibility to change** | None | Much/Great | Great |
| **Layout flexibility** | None | Possibly much | Possibly some |

**Table 1. Comparison of three translation approaches**

One example of community-based translation can be found in TidBITS [18], a technology newsletter. Teams of roughly five voluntary members work on the translation for one issue. One team member acts as coordinator to assign articles to the other four, and assemble the finished work. This administrative hierarchy has been found to work well, allowing an extra level of style and quality checking. The coordinator is in charge of maintaining the quality of the translation. Often, he/she also maintains a list of technical terms and their previous translations to help novice volunteers in the process. However, the initiative is slow and translations for most languages are behind at least a few issues. One of the co-authors of this paper was a volunteer to translate to TidBits to Spanish.

OmniWeb [12] is a browser created for Mac OSX and supports more languages (nineteen) than any other web browser for the Macintosh operating system. Run by The Omni Group, it is currently accepting help from volunteers to help translate the interface.

Another example is the Translation Project [19] which aims to translate free software packages into various languages. Its translators are voluntary community members who form independent teams for each language. Each team has its own e-mail address for contacting, and its own page with progress stats and member contacts. Members work together to translate on teams in different ways, but roles in the project include: maintainers, coordinators, and enthusiasts. A number of features are present to help these translators including a Robot which accepts submissions and does some error checking to be sure the translator has filed his/her disclaimer, a number of mailing lists for translation discussion, and a matrix which shows percentages of over ninety packages which have been translated into the forty-six different languages.

The 'Google in your language' beta translation initiative [5] is another example of community-based translation. Users of Google register to translate into one of one hundred and forty-eight languages, and are then presented with items (strings of words) which appear on Google's pages that could need translating or editing. Users use a system to type translations for these strings and instantly submit their changes for quick results. Users may also discuss concepts or problems involving the translation with other translators on community discussion boards.

There is a distinction between individual efforts and group efforts that needs to be drawn. In developer or administrator-based translations, there is one person or one small group of people who create translations and supply them for the users. It is a single result made by a single entity. With community-based translation, all members do small parts of the bigger picture on their own but their ability to collaborate with one another can keep these separate translations consistent in style and vocabulary. Harnessing the ability for users to collaborate online as a community has great potential for translation. See Table 1 for comparisons among the three different translation methods discussed here.

**MOTIVATION**

The motivation for our work comes from our research and development on CITIDEL. Initially we wanted to support the Hispanic community by providing a Spanish interface to CITIDEL. When the issue of sustainability was raised in our design meetings, we realized that had neither the resources nor the time to provide such translations. The result was to turn the translation effort into a community led effort. Our initial endeavor is to help produce a Spanish user interface for CITIDEL, however, there is nothing which prevents this process from being applied to any other language.

The initial focus for this translation project is on the Hispanic community, partly because of their fast growth in the United States. The 2000 U.S. Census [2] reported that Hispanics account for 12% (32.8 million) of the total population and Spanish is the second most spoken first language in the world. As the digital divide begins to decrease, the internet is drawing in more and more women and minorities into active roles [14]. Pew Studies [13] show that 50% of Hispanics in the U.S. who are 18 years old or older have used the Internet as of 2001; this is a 25% increase during the year 2000. This is in comparison to 58% of white adults and 43% of African-American adults who are online. Hispanics are typically very active online users with a variety of interests and uses. Therefore, it seems that this audience would be an extremely important and willing one for CITIDEL. Hispanics account for less than 3% of the population in computer science in the U.S. Spanish translations therefore might help bridge the language barrier and allow more Hispanics access to the educational resources within CITIDEL.

A secondary goal for the translation center is to increase the number of users that come to CITIDEL. This, we expect, will in turn increase the number of resources available in CITIDEL. If we are successful increasing the participation of non-English speakers, we might even attract educational resources produced in other languages. These would then become candidates for translation into English, thus increasing the number of resources in the system.

**EVALUATION OF APPROACHES**

As part of the of the analysis process for the design of our translation center we evaluated the language used in several translations. The evaluation was done to aid in our design, by having a principled way to evaluate the quality of the resulting language from the three approaches. In this section, we describe an evaluation of pages translated by developers, by community and by automated programs. The goal was to try to get an assessment of the quality of translation by each of the three approaches. Note, however, that this is not an exhaustive evaluation of all three approaches. For example, we used the AltaVista translation service, Babel Fish [1], which in turn uses the Systran software [17]. We did not, however, explore different versions of the software, nor the effect of configuration of the software. These should have an effect on the resulting translation.

We selected a few representative pages from different sites on the web and evaluated them based on a pre-defined rubric. The goal of the evaluation was to cover pages developed specifically in multiple languages, pages developed in English and translated by a community of volunteers and pages developed in one language (English or Spanish) and pages that were automatically translated.

The rubric, loosely based off the SUN Quality Assurance Document used to evaluate translations for their systems [16], covered four main areas of language use: structure, vocabulary, style and conveying the message. Each area has a number of subparts going from poor (0) to excellent (positive number) with a concise definition for each. The better the translation, the higher score the translation will get. Below are briefly presented the evaluation categories and the points for each (shown at the beginning of each line).

**Structure:**
0) Structures are those of the original language
1) Structures are acceptable in the language to which the document was translated, but calques are predominant from the original language
2) Structures are acceptable in the language to which the document was translated, with few calques from the original language
3) Structures are native to the language to which the document was translated

**Vocabulary: Cognates**
0) False cognates, direct translations, original language is very present
1) Use of cognates, with few false cognates, idiomatic expressions are translated directly
2) Few use of cognates, no false cognates, idiomatic expressions are well translate
3) Hardly any or no use of cognates, no direct translations

**Vocabulary: Meanings**
0) Words with more than one meaning do not receive an accurate translation
1) Words with more than one meaning receive an accurate translation

**Vocabulary: Spellings**
0) The document was obviously translated from the original language
1) It is not obvious that the document is a translation.

**Style: consistency**
0) Not consistent with the original language
1) Consistent with the original language

**Style: Punctuation, abbreviations**
0) Punctuation and abbreviations are consistent with the original language
1) Punctuation and abbreviations are accurate to the language to which the document was translated

**Message**
0) The document is not comprehensible, or requires a lot of interpretation to comprehend.
1) The ideas are stated, but needs some interpretation
2) The main points are present, but the ideas do not flow; choppy, disjointed
3) The ideas are clearly expressed

A perfect score would entail 13 points. We selected pages to evaluate that cover the translation approaches mentioned previously. Specifically we chose the following pages listed below.

**CITIDEL Pages**. We selected pages from CITIDEL because it represents our target application. The translation service is being developed particularly for CITIDEL. We selected two parts of CITIDEL, the text available on the home page that is used to welcome users to the site, and the text on the classification schemes used in CITIDEL. Both of these sections include large portions of text that would be prime candidates for translation, and they both include some amount of technical language.
The two pages evaluated were available on the web at the time of this writing:
Home page         http://www.citidel.org/
Classification page   http://www.citidel.org/?op=cbrowse

**Apple Mail Program Pages**. We selected the page that describes the Mail program available on the OS X for the Apple Macintosh computer because it is available in both English and Spanish. Also, this page is of a technical nature, it describes many of the features of the Mail program using technical language. The Spanish page of the Mail program is part of the Latin American site for Apple Computer, the whole site is an example of a web site redesigned in a different language, including having new icons and graphics where appropriate.
The two original pages were available on the web at the tmie of this writing:
*English*:  http://www.apple.com/macosx/jaguar/mail.html
*Spanish*:  http://www.apple.com/la/macosx/jaguar/mail.html

**TidBits.** TidBits [18] is an online newsletter that discussed Apple and Macintosh technology. It has been in publication since 1990. It features a volunteer-led effort to translate it to different languages. It is available in Spanish, Russian, Japanese, Portuguese, German, French, and Dutch. Each of the communities translates TidBits issues at their own pace, some are lagging other are very much active and current.

For this evaluation, we used the July 7th, 2002 edition of the newsletter because that was the last translation into Spanish. We evaluated only the story "Living under the Snow dome". The two pages used were available on the web at the time of this writing:
*English*:
http://www.tidbits.com/tb-issues/TidBITS-637.html
*Spanish*:
http://www.tidbits.com/tb-issues/lang/es/TidBITS-es-637.html

To some of these pages, we applied the Babel Fish translation service to obtain some machine-translated pages. The resulting mix of pages is shown on Table 2 below.

| Page | Language | Method | Score |
|---|---|---|---|
| 1. Citidel Home | English | Source | - |
| 2. Citidel Home | Spanish | Babel Fish | 1 |
| 3. Citidel Home | English | Babel Fish of 2. | 8 |
| 4. Citidel Classf. | English | Source | - |
| 5. Citidel Classf. | Spanish | Babel Fish | 1 |
| 6. Citidel Classf. | English | Babel Fish of 5. | 4 |
| 7. Apple Mail | English | Source | - |
| 8. Apple Mail | Spanish | Babel Fish | 2 |
| 9. Apple Mail LA | Spanish | Dev. Translation | 13 |
| 10. Apple Mail LA | English | Babel Fish | 3 |
| 11. TidBits #637 | English | Source | - |
| 12. TidBits #637 | Spanish | Community | 13 |
| 13. TidBits #637 | Spanish | Babel Fish | 1 |

Table 2. Pages used in the Evaluation

**RESULTS AND DISCUSSION**
The results of this evaluation were as expected: human translation and community translation produce higher quality translations than machine-translations for the selected pages and the software used. Table 2 shows the scores of the evaluation of each page.

None of the pages translated by Babel Fish received higher than a 2. It is worth noting, however, that the pages translated via Babel Fish to Spanish and then translated back to English produced higher quality translations (scores of 8 and 4) than the original translations. This is a side effect that the translation that Babel Fish is doing maintains much of the structure of the English language in the Spanish translation. The result of translating back into the source language is that the resulting structure is appropriate because is the structure from the source language, in this case English.

The pages translated by the community (TidBits) or by a development team (Apple's Mail page) received perfect scores. In the case of the Apple's page the translation is more completed than the machine-translation because they make use of images that contain text in them and these are not translated by Babel Fish. Furthermore, this site used some marketing slogans regarding the Mail's Junk Mail filter which would not translated properly. As a matter of fact, the Spanish ("Correo basura a la basura") and English ("The end of junk mail") versions of the slogans were very different, requiring human translation of the content.

From this simple evaluation, we learn that the translation services provided by one of the common services available on the web do not produce enough quality to trust the translation of the interface for CITIDEL.

**DESIGN OF CITIDEL TRANSLATION CENTER**
For CITIDEL, we are taking a voluntary community-based approach to translating. A 'translation community center' will be established to allow users to easily and accurately translate parts of the interface. Administrators would only have a basic part, in overseeing the running of the center and making any changes to the layouts of pages, if the community demands them.

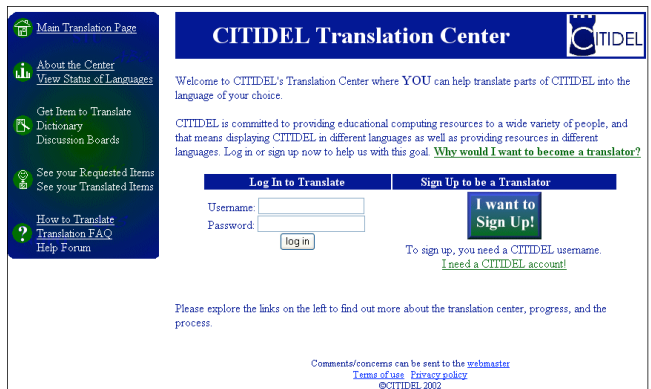

Figure 1. Designed front page of the Translation Center

Nielsen's research points to an important distinction- that of the difference between translated interfaces and localized interfaces. With the former, the pure language of the page is presented in another language, while the basic style, look, and feel of the page remain in their original spirit. Localized interfaces, however, can look drastically different, reflecting the specific cultures of the region of the world which speaks the language. For our project, we are concentrating on translated interfaces rather than localized ones.

There are still cultural problems to be dealt with, however. For example, there are differences in technical vocabulary used in Spanish in different countries around the world. For example, in Spain the computer is called "ordenador" while in Puerto Rico it is called "computadora". Would it be best to offer different versions of the site for Spanish (Puerto Rico), Spanish (Mexico), and Spanish (Spain)? If not, what will happen when there are disputes over specific terms to use? This is an issue this project hopes the community itself will be able to work out on its own. Tools for interaction and collaboration including language-specific discussion forums, a dictionary of computer terms with translations, and polls should help them come to a consensus as to what to do. Perhaps, in such disputes, the community will decide upon common terminology. Or perhaps they will decide different versions of the language are needed. In the end, it is an issue to be resolved by the translators and users of the translations themselves.

In our designed self-regulating task-based community, users will be given items (currently single words or strings of words) to translate and many features which hope to assist the translators in their tasks. These currently include:

- Creating Translations
  o List of items that need to be translated (shown in Figure 3)
  o Ability for users to enter translations of the interface (shown in Figure 4)
  o Way for users to edit translations

- Basic Features
  o Meters/percentages as to how much has been translated into particular languages
  o How-to translate section/tutorial
  o Frequently-Asked-Questions List
  o Personal binder containing items translated or requested
  o Way to recommend a page/section to be translated

- Community/Interactive Features
  o Dictionary of field (computer/technology) terms with their translations and definitions
  o List of translators who wish to be publicly contacted and their contact information
    o Way to rate and review translations
    o Suggestion forum- direct feedback and improvement by users
    o Help forum- users assist each other in translating
    o General forums- general discussion about translation
    o Language-specific forums- for language-specific inquiries
    o Polls to allow community members to decide on disputed issues

Many of these features are intended to help the translation process while others allow users to interact as a community on issues of translation.

**How to Translate**
Similar to Google (Figure 2), our system (Figure 3) will have a listing of items which need to be translated. In the future, a priority system developed based upon where the item occurs, how frequently it is seen, number of members who requested it be translated, and reviews made on quality of the translation will help in the ordering of this list. Translators may choose from items on this list to review, or request a single "random" translation and bypass the task of looking through the list and choosing.

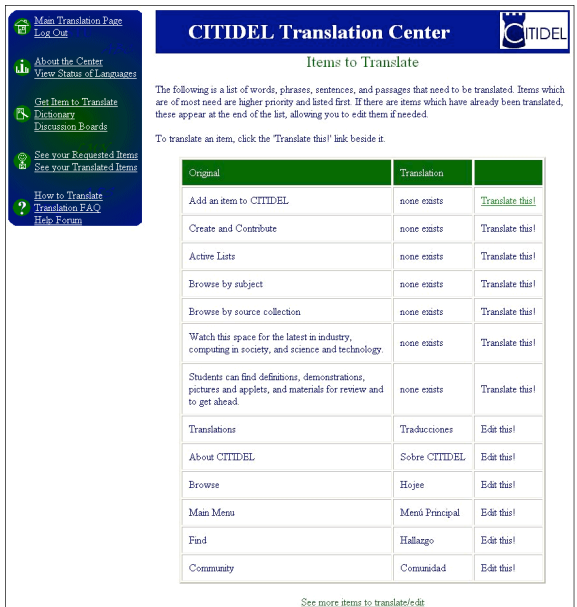

Figure 3. Our listing of items that can be translated or edited into another language

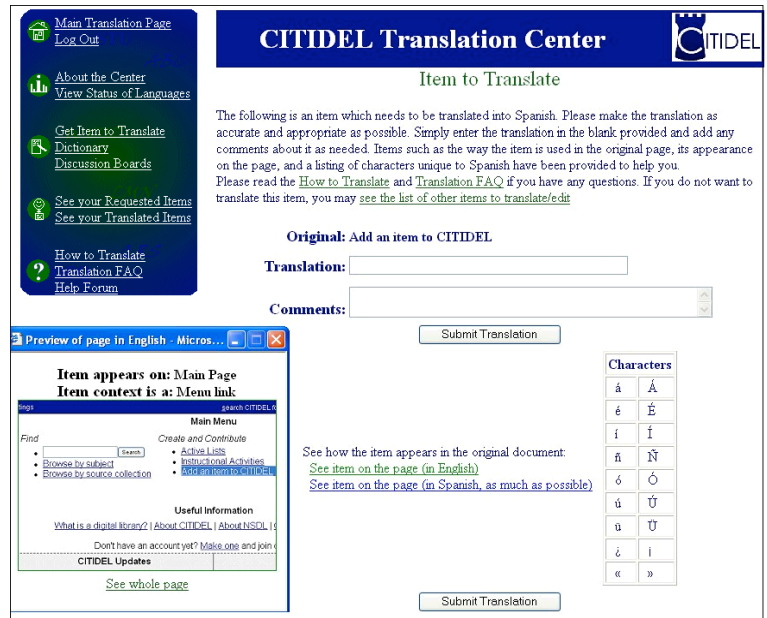

Figure 4. Our page for translating one item. Features include: links to help tools, comment section for translation-specific concerns/comments, table of language-specific characters, portion of page in English or Spanish where the item appears.

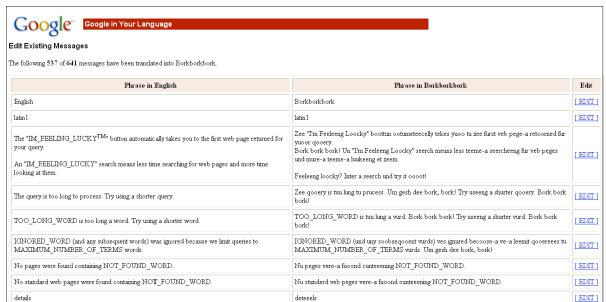

Figure 2. Google's listing of items (messages) which can be edited into another language

With Google's popular languages, only trusted and authorized translators are allowed to translate items for the first time, while ordinary translators may just edit existing items. This sometimes creates bottlenecks and slows translation. It is hoped that our added features and review system will help translators create quality translations thus not requiring the extra layer of trusted translators. This community review aspect takes the place of editors, assemblers or moderators which most of the community-driven systems we discussed in the background section operate with. Instead, we will let the community as a whole decide what is a quality translation, what industry terms to use, and what issues are of concern to their particular translations.

Unlike Google, however, many features are provided in our system for translators when they attempt to translate an item, as seen in Figure 4. Easy links to help resources such as a 'how to translate tutorial' and a 'translation FAQ' are provided in multiple locations on the page. Translators are able to easily copy-and-paste characters specific to their language into their translation without switching screens, keyboard settings or needing to know codes. They are also able to see the context in which the item they are translating occurs. They can see a portion of the original page in English or in the translated language (in as much as it is translated) with their item highlighted and with another optional link to see the entire page. The translator is given further perspective with a listing of the page it originated on and the category or type of item, such as 'menu link' or 'informational text'. The translator is also able to enter comments about the translation which will become public for all translators and perhaps appear automatically on the language's working forum for discussion. These comments could refer to possible difficulties the translator had with the translation, or to culture-specific terminology for examples. When editing items, translators see all of the associated comments and discussion submitted with the item, which should help them understand issues or complications with it as they attempt to edit.

**Basic Features**

There are a number of basic, less interactive features which will help translators with their task. Meters will be displayed on several pages, including the main page, showing how much has been translated into particular languages. This should help translators feel they are working towards and goal and making a difference as the percentages go up. Help documents such as a 'how to' tutorial which stresses maintaining consistent style and vocabulary throughout the site and an FAQ will help with the basic task of translating and quickly answer questions. A help forum will exist for discussion and questions not covered in these two documents. Each translator will have their own section or a binder of sorts. Items they have translated will be placed in this personal binder so they may revisit them any time to see reviews or edit them. Users may request certain important pages or parts of pages to be translated. These pages would have an extra level of priority, and allow users of CITIDEL who do not speak English to voice their opinions about the usefulness of translated items. Translators may also request that certain parts of CITIDEL be translated, and these items of request will appear in a translator's binder as well with a notice or indication when one is translated.

**Community Features**

There are also a number of features which are interactive and support the sense of community and community task. An optional list allows translators to display their contact information as well as how many items they have translated, as well as contact other translators. Translators may wish to meet each other in person to discuss translation or language-related concerns, or simply discuss the areas of computing that CITIDEL covers. At least four different kinds of forums will allow discussion among the translators. The forums include a general translation forum, a help forum, a suggestion forum, and language forums where translators may discuss (in whatever language they prefer) issues which are related to their language of choice. Forums provide ways of inquiry and solution based on other community members. To ensure consistent use of industry terms, a dictionary of computer terms (including definitions and threaded comments from users) can be added to. Translators as a whole or in sub-communities for each language may create and answer polls as needed, relating to translation issues. Forum discussions and polls may help resolve differences or disputes, should any arise. While the translators are in control of content, a user-review system helps check that translations appropriate on the basis of content, quality and style. Translations can be rated and reviews be posted with comments. These ratings and reviews may be helpful in the editing process. These features will hopefully allow the community to not only regulate themselves but produce translations of agreeably good quality and similar style.

## CONCLUSION

In this paper we have presented the design of an online community-based translation center to support translation of the interface of CITIDEL. We provided a brief evaluation of the language produced by three typical translation efforts: developer-based, community-lead, and machine (automated) translation. The software design presented here is currently being implemented and will be in use by volunteers around Virginia Tech and around the world. We expect to conduct further evaluations of the resulting interface once the system is running. We are particularly interested in observing the community dynamics and community-imposed quality controls that result from this translation center.